\documentclass[leqno,12pt]{article}
\usepackage{secdot}%section number+dot
\usepackage{sectsty}%sectiontitle
\usepackage{extarrows,color}
\usepackage{authblk}%author
\usepackage{verbatim}
\allowdisplaybreaks[3]%equation over pages
\usepackage{bbm}
\usepackage[top=1.25in, bottom=1.25in, left=1.25in, right=1.25in]{geometry}
\usepackage{mdwlist}
\usepackage{amsmath}

\usepackage[symbol]{footmisc}
\DefineFNsymbols{nu}{* 1 2}\setfnsymbol{nu}   % author number
\title{Stochastic maximum principle under probability distortion \thanks{This research is supported by Macao Science and Technology Development Fund FDCT 025/2016/A1 and Southern University of Science and Technology Start up fund Y01286120.}}
\author{Qizhu Liang \thanks{Department of Mathematics, University of Macau, Macau, China (yb47422@umac.mo).} ~and~ Jie Xiong  %\samethanks[2]
 \thanks{Department of Mathematics, Southern University of Science and Technology, Shenzhen, China (xiongj@sustc.edu.cn).}}
\date{}

\begin{document}
\newtheorem{theorem}{Theorem}[section]
\newtheorem{definition}[theorem]{Definition}
\newtheorem{corollary}[theorem]{Corollary}
\newtheorem{lemma}[theorem]{Lemma}
\newtheorem{assumption}[theorem]{Assumption}
\newtheorem{example}[theorem]{Example}
\newtheorem{remark}[theorem]{Remark}
\newcommand{\qed}{\hfill\rule{2mm}{3mm}\vspace{4mm}}

\numberwithin{equation}{section}

\def\ep{{\varepsilon}}\def\si{{\sigma}}\def\la{{\lambda}}

\def\({\left(}\def\){\right)}

%\newfam\msbmfam\font\tenmsbm=msbm10\textfont
%\msbmfam=\tenmsbm\font\sevenmsbm=msbm7
%\scriptfont\msbmfam=\sevenmsbm\def\bb#1{{\fam\msbmfam #1}}
%\def\AA{\bb A}\def\BB{\bb B}\def\CC{\bb C}
%\def\mathbbm{E}{\bb E}\def\HH{\bb H}\def\KK{\bb K}
%\def\LL{\bb L}\def\NN{\bb N}\def\mathbbm{P}{\bb P}\def\QQ{\bb Q}
%\def\mathbbm{R}{\bb R}\def\SS{\bb S}\def\TT{\bb T}\def\ZZ{\bb Z}

\def\EE{\mathbbm E}
\def\PP{\mathbbm P}
\def\RR{\mathbbm R}
\def\cF{\mathcal{F}}
\def\cU{\mathcal{U}}

\maketitle

\begin{abstract}
Within the framework of the cumulative prospective theory of Kahneman and Tversky, this paper considers a  continuous-time behavioral portfolio selection problem whose model includes both running and terminal terms in the objective functional. Despite the existence of $S$-shaped utility functions and probability distortions, a necessary condition for the optimality is derived. The results are applied to various examples.

% \begin{flushleft}
\vspace{.5cm}\noindent
{\bf Key words:}
cumulative prospective theory,
$S$-shaped utility function,
probability distortion,
stochastic maximum principle,
behavioral portfolio optimization
%\end{flushleft}

\vspace{.5cm}\noindent
{\bf AMS subject  classifications}. Primary 93E20; Secondary 91G80.

\end{abstract}

%\sectionfont{\centering\textmd}
\section{Introduction}
Expected utility theory (EUT) prevailed for a long time as the dominant preference measure under risk. Along with the theory in continuous financial portfolio selection problems, many approaches, such as dynamic programming, stochastic maximum principle, martingale and convex duality have been developed, see Merton \cite{Merton 1969}, Peng \cite{Peng 1990}, Duffie and Epstein \cite{Duffie and Epstein 1992}, Karatzas et al. \cite{Karatzas et al. 1991}.
The EUT, proposed by von Neumann and Morgenstern \cite{von Neumann and Morgenstern 1944},   is premised on the tenets that the utilities of outcomes are weighted by their probabilities and decision makers are consistently risk averse. These, however, have been violated by substantial phenomena.

Allais  \cite{Allais 1953} argues that individuals evaluate (overweight or underweight the probability of) every outcome depending on the other outcomes of a prospect via a paradox. Related studies in response to this fact are Fishburn \cite{Fishburn 1988}, Schmeidler \cite{Schmeidler 1989}, etc. On the other hand, risk-seeking behavior pervades decision problems, e.g., people would love to spend $x$ on the lottery with expected payoff no more than $x$. Likewise in loss situation, people usually prefer a possible large loss to a certain loss. Quite a few economists, such as Yaari \cite{Yaari 1987}, have investigated the modification of EUT on these challenges.

The most notable effort to alternate EUT is the prospect theory (PT) of Kahneman and Tversky \cite{Kahneman and Tversky 1979}, which takes investors' psychology into account in the face of uncertainty. Later the PT was evolved into cumulative prospect theory (CPT)  by Tversky and Kahneman \cite{Tversky and Kahneman 1992}. A significant difference between CPT and PT is that weighting is applied to the cumulative distribution functions, but not applied to the probabilities of individual outcomes; that is, the new version can be extended to the continuous distributions. The key elements of CPT include i)
A benchmark serves as a base point to distinguish gains from losses. Without loss of generality, it is assumed to be $0$ in this paper.
ii)
Utility functions are concave for gains and convex for losses, and steeper for losses than for gains.
iii)
Probability distortions (or weighting) are nonlinear transformation of the probability measures, which overweight small probabilities and underweight moderate and high probabilities.

There have been burgeoning research focuses on merging the CPT or PT into portfolio choice issues. Most of them are limited to the discrete-time setting, see for example Benartzi and Thaler \cite{Benartzi and Thaler 1995}, Shefrin and Statman \cite{Shefrin and Statman 2000}, Levy and Levy \cite{Levy and Levy 2003}. The pioneering analytical research on continuous-time asset allocation featuring behavioral criteria is done by Jin and Zhou \cite{Jin and Zhou 2008}. Since then, a few extensive works have been published, see He and Zhou (\cite{He and Zhou Feb 2011}, \cite{He and Zhou Apr 2011}), and Jin and Zhou \cite{Jin and Zhou 2013}.
Jin and Zhou \cite{Jin and Zhou 2008} developed a new theory to work out the optimal terminal value in continuous-time CPT models, featuring both $S$-shaped utility functions and probability distortions.
Their prominent idea is to change the decision variable from the random variable to its quantile function, such that the non-concave/convex objective turns to be a concave functional.
The whole machinery is quite involved.
To achieve the optimal control process that replicates the optimal terminal value, a further calculation is necessary.
Nonetheless, their theory aims at a particular portfolio choice problem in a self-financing market (i.e. there is no consumption or income).

The main motivation of our work is to deal with probability distortion for model with consumption. In order to come closer to reality, bankruptcy is not allowed in our problem.
Below are two examples which motivate our work.

Let $T>0$ be a fixed time horizon and $(\Omega,\mathcal{F},\mathbbm{P},\{\mathcal{F}_t\}_{t\ge 0})$ a filtered complete probability space on which is defined a standard $\mathcal{F}_t$-adapted $m$-dimensional Brownian motion $W_t\equiv(W^1_t,\cdots,W^m_t)^\top$ with $W_0=0$. It is assumed that $\mathcal{F}_t=\sigma\{W_s:0\le s\le t\}$, augmented by all the null sets. Throughout this paper $A^\top$ denotes the transpose of a matrix $A$; $a^\pm$ denote the positive and negative parts of the real number $a$.

Let $(\widetilde{\Omega},\widetilde{\mathcal{F}},\widetilde{\mathbbm{P}})$ be a copy of the probability space $(\Omega, \mathcal{F}, \mathbbm{P})$. For any random variable $\xi$ over $(\Omega, \mathcal{F}, \mathbbm{P})$ we denote by $\widetilde{\xi}$ a copy of $\xi$ defined on $(\widetilde{\Omega},\widetilde{\mathcal{F}},\widetilde{\mathbbm{P}})$. The expectation $\widetilde{\mathbbm{E}}[\cdot]=\int_{\widetilde{\Omega}}(\cdot)d\widetilde{\mathbbm{P}}$ acts on the variable $\widetilde{\omega}\in\widetilde{\Omega}$ only. In what follows, we replace $F_{\tilde{Y}}(Y)=\widetilde{\mathbbm{P}}\{\widetilde{Y}\le Y\}$ by $F_Y(Y)$ for convenience.

\begin{example}
$\textbf{(Investment vs. Consumption)}$
We illustrate a model from Pham (\cite{2009}, Section 3.6.2). The financial market consists of a bond with price $S^0_t$ given by
\[dS^0_t=r_tS^0_tdt,~~S^0_0=s_0>0,\]
and $m$ stocks with prices per share $S^i_t,~i=1,\cdots,m$, modeled by the geometric Brownian motions given by
\[dS^i_t=S^i_t\Big(b^i_tdt+\sum^{m}_{j=1}\sigma^{ij}_t dW^j_t\Big),~~S^i_0=s_i>0.\]
The interest rate $r_t$, the vector $b_t=(b^1_t,\cdots,b^m_t)^\top$ of stock appreciation rates, and the volatility matrix $\sigma_t=\{\sigma^{ij}_t\}_{1\le i,j\le m}$ are taken to be $\mathcal{F}_t$-progressively measurable stochastic processes.

In this financial market, bankruptcy is not allowed. The wealth process $X.$ is required to be positive. Let $u^i_t$ (which may be negative, or may exceed 1) be the proportion of wealth invested in stock i, and $c_t$ be the consumption per unit of wealth at time $t$. The remaining proportion $1-\sum^{m}_{i=1}u^i_t$ is invested in the bond.  Then $X_t$ evolves according to the forward stochastic differential equation (SDE)
\[
\left\{
\begin{array}{ccl}
dX_t&=&\sum^{m}_{i=1} \frac{u^i_tX_t}{S^i_t}dS^i_t+\frac{(1-\sum^m_{i=1}u^i_t)X_t}{S^0_t}dS^0_t-c_tX_tdt\\

&=&X_t(r_t+(b_t-r_t\mathbf{1_m})^\top_tu_t-c_t)dt+X_tu^\top_t\sigma_tdW_t,~~~t\in[0,T];\\
X_0&=&x_0>0,
\end{array}
\right.
\]
where $u_t=(u^1_t,\cdots,u^m_t)^\top$ and $c_t$ together is the portfolios of the investor. Like in most papers in the literature, we define a trading strategy or portfolio as the proportions or fractions of wealth allocated to different assets, see Merton \cite{Merton 1969}, Karatzas et al. \cite{Karatzas et al. 1991}, Karatzas and Shreve \cite{Karatzas and Shreve 1998}.

Within the continuous-time CPT framework of Jin and Zhou \cite{Jin and Zhou 2008}, the objective is to find the optimal consumption path $c_\cdot$ and the portfolio strategy on shares $u_\cdot$ such that the prospective preference
\[
J(c_\cdot,u_\cdot)=\int_0^T\int^{\infty}_0\mathbbm{P}\{\zeta(c_tX_t)>y\}dydt
+\int^{\infty}_0w(\mathbbm{P}\{l(X_T)>x\})dx.
\]
achieves the maximum. Here $\zeta(\cdot),l(\cdot):\mathbbm{R}^+\to\mathbbm{R}^+$ are the investor's utility functions for consumption and terminal wealth, respectively, and $w(\cdot):[0,1]\to[0,1]$ represents the distortion of probability. There is no distortion on consumption. In fact, the prospective functional could be written as
\[
J(c_\cdot,u_\cdot)=\mathbbm{E}\int^T_0\zeta(c_tX_t)dt
+\mathbbm{E}\big(l(X_T) w'(1-F_{X_T}(X_T))\big).
\]
\end{example}

\begin{example}
$\textbf{(Investment vs. Gambling)}$
In addition to the investment in aforementioned market, an investor is allowed to buy lottery tickets.  Here the wealth is required to be positive as well.
For simplicity, let $c_t\in\mathbbm{R}^+$ be the wager per unit of wealth at time $t$ and $K_t$ be the odds of winning.
For instance, if $K_t$ is $8$ with probability $0.1$ and $-1$ with probability $0.9$, the investor will win $8c_tX_t$ with probability $0.1$ and lose the wager $c_tX_t$ with probability $0.9$ at $t$. The wealth process is governed by
\[
\left\{
\begin{array}{ccl}
dX_t&=&X_t(r_t+(b_t-r_t\mathbf{1_m})^\top_tu_t)dt+X_tu^\top_t\sigma_tdW_t+K_t c_tX_tdt,~~~t\in[0,T];\\
X_0&=&x_0>0,
\end{array}
\right.
\]
where $u_t=(u^1_t,\cdots,u^m_t)^\top$ and $c_t$ consist of the portfolio of the investor. For this case, the portfolio selection problem is to find the most preferable portfolio to maximize the distorted expected payoff
\[
\left.
\begin{array}{l}
J(c_\cdot,u_\cdot)=\int_0^T\Big(\int^{\infty}_0\varpi_+(\mathbbm{P}\{\zeta_+(K_t^+c_tX_t)>y\})dy\\
~~~~~~~~~~~~-\int^{\infty}_0\varpi_-(\mathbbm{P}\{\zeta_-(K_t^-c_tX_t)>y\})dy\Big)dt
+\int^{\infty}_0w(\mathbbm{P}\{l(X_T)>x\})dx,
\end{array}
\right.
\]
\iffalse
\begin{align*}
J(c_\cdot,u_\cdot)=&\int_0^T\Big(\int^{\infty}_0\varpi_+(\mathbbm{P}\{\zeta_+(K_t^+c_tX_t)>y\})dy\\
&-\int^{\infty}_0\varpi_-(\mathbbm{P}\{\zeta_-(K_t^-c_tX_t)>y\})dy\Big)dt
+\int^{\infty}_0w(\mathbbm{P}\{l(X_T)>x\})dx,
\end{align*}
\fi
where $\zeta_+(\cdot),\zeta_-(\cdot):\mathbbm{R}^+\to\mathbbm{R}^+$ are utility functions measuring the gains and losses of gambling, respectively.  $\varpi_+(\cdot),\varpi_-(\cdot):[0,1]\to[0,1]$ represent the distortions in probability for the gains and losses, respectively. $w(\cdot)$ and $l(\cdot)$ are as same as those in the last example.
Straightforwardly, the distorted payoff could be written as
\[
\left.
\begin{array}{l}
J(c_\cdot,u_\cdot)=
\mathbbm{E}\int_0^T\Big(\zeta_+(K_t^+c_tX_t)\varpi'_+\big(1-F_{K_t^+c_tX_t}(K_t^+c_tX_t)\big)\\
~~~~~~~~~~~~~~~~~~~~~-\zeta_-(K_t^-c_tX_t)\varpi'_-\big(1-F_{K_t^-c_tX_t}(K_t^-c_tX_t)\big)\Big)dt\\
~~~~~~~~~~~~+\mathbbm{E} \big(l(X_T)w'(1-F_{X_T}(X_T))\big).
\end{array}
\right.
\]
The objective is to find an optimal portfolio $(u.,c.)$ to maximize $J$.
\end{example}

 In general, we will consider optimization problems with probability distortions and running utilities.
Resulting from the distorted probability, time-consistency of the conditional expectation with respect to a filtration is invalid. Thus the dynamic programming approach is failed upon the underlying problem.  On the other hand, the quantile formulation introduced in Jin and Zhou \cite{Jin and Zhou 2008} is feasible to those of the control being a random variable rather than a stochastic process. It doesn't work on the running terms. In this paper, we therefore employ the stochastic maximum principle to conquer the aforementioned difficulties, and strive to acquire the necessary conditions of the optimal control process for the general optimization problems.

The rest of this article is organized as follows. Next section will formulate a general continuous-time portfolio selection model under the CPT, featuring $S$-shaped utility functions and probability distortions. After that, the main results of this paper are presented. The stochastic maximum principle is used to obtain the necessary conditions for optimality in Section 3. In Section 4, we apply our general result to three interesting examples. Final concluding remarks are presented in the last section.

\section{Problem Formulation and Main Result}
We define a positive state process
\begin{equation}\label{state}
\left\{
\begin{array}{ccl}
 dX_t  &=&   b(t,u_t,X_t)dt+\sigma(t,u_t,X_t) dW_t  \\
  X_0  &=&  x_0>0,
\end{array}
\right.
\end{equation}
and the agent's prospective functional
\begin{equation}\label{eqJ0}
\left.
\begin{array}{l}
J(u_\cdot)=\mathbbm{E}\int^T_0 \big(\zeta_+(u_t^+)\varpi'_+\big(1-F_{u^+_t}(u^+_t)\big)
-\zeta_-(u^-_t)\varpi'_-\big(1-F_{u^-_t}(u^-_t)\big)\big)dt\\
~~~~~~~~~~~
+\mathbbm{E}\(l(X_T)w'\big(1-F_{X_T}(X_T)\big)\),
\end{array}
\right.
\end{equation}
where $W_t$ is a 1-dimensional Brownian motion, $u_\cdot$ is a control process taking values in a convex set $U\subseteq \mathbbm{R}$.  

According to CPT, the following assumptions will be in force throughout this paper, where $x$ denotes the state variable, and $u$ denotes the control variable.
\begin{itemize}
\item[(H.1)] $b(\cdot, \cdot,\cdot):[0,T]\times U\times\mathbbm{R}^+\to\mathbbm{R}$, $\sigma(\cdot, \cdot,\cdot):[0,T]\times U\times\mathbbm{R}^+\to\mathbbm{R}$, are continuously differentiable with respect to $(u,x)$.
The first derivatives of $b, \sigma$ with respect to $(x,u)$ are Lipschitz continuous. We further assume $b(t,u,0)=\si(t,u,0)=0$.
\item[(H.2)]
$\zeta_\pm(\cdot),l(\cdot):\mathbbm{R}^+\to\mathbbm{R}^+$ are supposed to be differentiable, strictly increasing, strictly concave, and satisfy $\zeta_\pm(0)=l(0)=0$ and the Inada conditions $\zeta'_\pm(0+)=l'(0+)=\infty$.
\item[(H.3)]
$\varpi_\pm(\cdot),w(\cdot):[0,1]\to[0,1]$, are differentiable and strictly increasing, with $\varpi_{\pm}(0)=w(0)=0$, $\varpi_{\pm}(1)=w(1)=1$. Moreover, the first derivatives of $\varpi_\pm(\cdot),w(\cdot)$ are all bounded.
\end{itemize}

A typical example of the utility function is $l(x)=\frac{x^\gamma}{\gamma},~0<\gamma< 1$, while that for the distortion function is the decumulative weighting function used in Lopes's SP/A theory \cite{Lopes 1987} which takes the form:
$
w(p)=\nu p^{\alpha+1}+(1-\nu)[1-(1-p)^{\beta+1}],
$
where $0\le\nu\le 1$ and $\alpha,\beta\ge 0$. Clearly, $p^{\alpha+1}$ and $1-(1-p)^{\beta+1}$ are convex and concave functions, respectively.
Define
\[
\mathcal{U}=\Big\{u: [0,T]\times\Omega\to U~|~u_t\mbox{ is }\mathcal{F}_t\mbox{-adapted and }\mathbbm{E}\int^T_0|u_t|^4dt<\infty\Big\}.
%\mathcal{U}=\{u: [0,T]\times\Omega\to U~|~u_t\mbox{ is }\mathcal{F}_t\mbox{-adapted and bound}\}.
\]

\begin{definition}\label{adm}
A control process $u_\cdot\in\mathcal{U}$ is said to be admissible, and $(u.,X.)$ is called an admissible pair, if
\begin{enumerate}
  \item[(1)] $X.$ is the unique solution of equation $(\ref{state})$ under $u.$;

  \item[(2)] both $u_\cdot^+$ and $u_\cdot^-$  possess continuous (except at 0) distribution functions;

  \item[(3)]
$\mathbbm{E}\int^T_0\Big(\big| \zeta_+(u_t^+)\varpi'_+\big(1-F_{u^+_t}(u^+_t)\big) \big|^8
+\big|\zeta_-(u^-_t)\varpi'_-\big(1-F_{u^-_t}(u^-_t)\big)\big|^8\Big)dt<\infty$.

\item[(4)]
$\mathbbm{E}\int^T_0\Big( \big|\frac{d}{du}\ln \zeta_+(u_t^+)\big|^8
  +\big|\frac{d}{du}\ln \zeta_-(u_t^-)\big|^8
  +\big|\zeta''_+(u_t^+)\big|^4
 + \big|\zeta''_-(u_t^-)\big|^4\Big)dt<\infty$.
\end{enumerate}
The set of all admissible controls is denoted by $\mathcal{U}_{ad}$.
\end{definition}

\begin{remark}
If $\zeta_\pm(u)=\frac{u^\gamma}{\gamma},0<\gamma<1$, the condition (4) is satisfied provided that the admissible control are restrict to those with $\EE \int^T_0|u_t|^{-8}dt<\infty$.
\end{remark}

Meanwhile, some technical assumptions for the terminal state are in force throughout this paper.
\begin{assumption}\label{ass}
The terminal state $X_T$ corresponding to the control process $u.\in\cU_{ad}$  has continuous distribution function, and \begin{equation}\label{assX}
  \EE\big| l(X_T)w'\big(1-F_{X_T}(X_T)\big)\big|^8
  + \EE \Big|\frac{d}{dx}\ln l(X_T)\Big|^8
  +\EE \big|l''(X_T)\big|^4 <\infty.
\end{equation}

\end{assumption}
The condition (3) in Definition $\ref{adm}$ as well as the first term of inequality $(\ref{assX})$ guarantee that the prospective functional $J(u_\cdot)$ is always finite. Generally, in the case that the supremum of $J$ is finite with bounded initial investment $x_0$, the model is regarded as well-posed; otherwise, it is ill-posed.

\begin{remark}\label{rem2.4}
If $l(x)=\frac{x^\gamma}{\gamma},0<\gamma<1$, $\hat{b}(t,u,x)=x^{-1}b(t,u,x)$ and $\hat{\si}(t,u,x)=x^{-1}\si(t,u,x)$ are bounded on $[0,T]\times\mathcal{U}\times\RR^+$, then $\EE \big|\frac{d}{dx}\ln l(X_T)\big|^8$ and $\mathbbm{E} \big|l''(X_T)\big|^4$ are finite. 

In fact, applying It\^o's formula to $X_t^{4\gamma-8}$, we finally get
\[
\left.
\begin{array}{ccl}
\mathbbm{E} \big(l''(X_T)^4\big)
&=&
(\gamma-1)^4X_0^{4\gamma-8}\cdot
\mathbbm{E}\exp\big\{(4\gamma-8)\big(\int^T_0(\hat{b}(t,u_t,X_t)\\
&&~~~~~~~~~~~~~~~
-\frac{1}{2}\hat{\sigma}^2(t,u_t,X_t))dt
+\int^T_0\hat{\sigma}(t,u_t,X_t) dW_t\big)\big\},
\end{array}
\right.
\]
which is bounded. So is $\EE \big|\frac{d}{dx}\ln l(X_T)\big|^8$.
\end{remark}

\begin{remark}
The continuity assumption of the distribution is not very restrictive due to the continuous model we  study. Here are some cases in which this condition is satisfied:\\
i) If the control is in Markovian feedback form, namely, $u_t=g(t,X_t)$ for a suitable measurable function $g$, the existence of the density $v(t,x)$ for the random variable $X_t$ follows from the general PDE theory because it satisfies the Chapman-Kolmogorov equation.\\
ii) Let $\hat{b}$ and $\hat{\si}$ be given in Remark \ref{rem2.4}. If $\hat{b}-\frac12\hat{\si}^2$ is bounded and $\hat{\si}^2$ is bounded below away from 0, then $X_T$ has a density.
\end{remark}
{\bf Proof:} For i) we refer the readers to Kusuoka and Stroock \cite{Kusuoka and Stroock 1984}, Kusuoka and Stroock \cite{Kusuoka and Stroock 1985}, Bouleau and Hirsch \cite{Bouleau and Hirsch 1991}, Kusuoka \cite{Kusuoka 2009} for many sufficient conditions for the existence of the density. Now we give a proof of ii).

Denote $c_t=\hat{\si}(t,u_t,X_t)$ and $Y_t=\ln X_t$. By It\^o's formula, we have
\[dY_t=\(\hat{b}-\frac12\hat{\si}^2\)(t,u_t,X_t)dt+c_tdW_t.\]
By Girsanov's theorem, there exists an equivalent probability measure $\tilde{P}$ and a $\tilde{P}$-Brownian motion $\tilde{W}$ such that
\[dY_t=c_td\tilde{W}_t.\]
Now we only need to prove that $Y_t$ has a density. For simplicity of notation, we assume that $Y_0=0$. Denote $F(\la)=\tilde{\EE}e^{i\la Y_T}$. Then,
\begin{eqnarray*}
F(\la)
&=&\lim_{n\to\infty}\tilde{\EE}\exp\(i\la\sum^{n-1}_{j=0}c_{t_j}\(\tilde{W}_{t_{j+1}}-\tilde{W}_{t_j}\)\)\\
&=&\lim_{n\to\infty}\tilde{\EE}\exp\(i\la\sum^{n-2}_{j=0}c_{t_j}\(\tilde{W}_{t_{j+1}}-\tilde{W}_{t_j}\)\)
\exp\(-\frac{\la^2}2 c^2_{t_{n-1}}(t_n-t_{n-1})\)\\
&\le&\lim_{n\to\infty}\tilde{\EE}\exp\(i\la\sum^{n-2}_{j=0}c_{t_j}\(\tilde{W}_{t_{j+1}}-\tilde{W}_{t_j}\)\)
\exp\(-\frac12 \ep_0\la^2(t_n-t_{n-1})\)\\
&\le&\cdots\\
&\le&\exp\(-\frac{\ep_0T}{2}\la^2\),
\end{eqnarray*}
where $0=t_0<t_1<\cdots<t_n=T$ is a partition of $[0,T]$ with $\max_j|t_{j+1}-t_j|\to 0$.

Since $F$ is in $L^1$, it is the Fourier transform of an $L^1$ function, which is the density of $X_T$.
\qed

Now, we are ready to state our problem and to present our main result.

$\textbf{Problem.}$ Our optimal control problem is to find $\bar{u}_\cdot\in\mathcal{U}_{ad}$ such that
\begin{equation}\label{pro}
J(\bar{u}_\cdot)=\max_{u_\cdot\in\mathcal{U}_{ad}}J(u_\cdot).
\end{equation}

Let $(\bar{u}_\cdot,\bar{X}_\cdot)$ be an optimal pair of the problem $(\ref{pro})$. Before stating the main result of this paper, we formulate the adjoint equation
\begin{equation}\label{adjoint}
\left\{
\begin{array}{ccl}
dp_t&=&-\big(b_x(t,\bar{u}_t,\bar{X}_t)p_t+\sigma_x(t,\bar{u}_t,\bar{X}_t) q_t
\big)dt+q_tdW_t,\\
p_T&=&l'(\bar{X}_T)w'(1-F_{\bar{X}_T}(\bar{X}_T)),
\end{array}
\right.
\end{equation}
where $b_x$ and $\si_x$ denote the partial derivatives (in $x$) of $b$ and $\si$, respectively.

\begin{theorem}\label{thm}
If $\bar{u}_\cdot$ is an optimal control with the state trajectory $\bar{X}_\cdot$, then there exists a pair $(p_\cdot,q_\cdot)$ of adapted processes which satisfies $(\ref{adjoint})$ such that a.e. $t\in[0,T]$,
\begin{equation}\label{smp}
p_tb_u(t,\bar{u}_t,\bar{X}_t)+\sigma_u(t,\bar{u}_t,\bar{X}_t)q_t
=\left\{
\begin{array}{ll}
-\zeta'_+\big(\bar{u}^+_t\big)
\varpi'_+\big(1-F_{\bar{u}^+_t}
(\bar{u}^+_t)\big)
&\mbox{ if }\bar{u}_t>0,\\
-\zeta'_-\big(\bar{u}^-_t\big)
\varpi'_-\big(1-F_{\bar{u}^-_t}
(\bar{u}^-_t)\big)
&\mbox{ if }\bar{u}_t<0,
 \end{array}
\right. a.s..
\end{equation}
\end{theorem}

\begin{remark}
Theorem $\ref{thm}$ remains true when $\zeta_\pm(\cdot),~l(\cdot)$ are replaced by functions which are twice continuously differentiable and take zero value at zero. 

Additionally, the model can be generalized. For instance, we may use $u_t^\pm X_t$ instead of $u_t^\pm$ in the objective functional.
Again, the risk preference can be defined as
\begin{align*}
J(u_\cdot)=&\mathbbm{E}\int^T_0 \Big(f(t,u_t,X_t)
+\zeta_+(u_t^+)
\varpi'_+\big(1-F_{u_t^+}(u_t^+)\big) \\
&~~-\zeta_-(u_t^-)
\varpi'_-\big(1-F_{u_t^-}(u_t^-)\big)\Big)dt
+\mathbbm{E}\(l(X_T)w'\big(1-F_{X_T}(X_T)\big)\),
\end{align*}
where $f(\cdot,\cdot,\cdot): [0,T]\times\RR^m\times\RR^+\to \RR$ is supposed to be twice continuously differentiable with respect to $u$ and $x$. 

The necessity of optimality for such a problem is
\begin{align*}
&p_tb_u(t,\bar{u}_t,\bar{X}_t)+\sigma_u(t,\bar{u}_t,\bar{X}_t)q_t
+\partial_u f(t,\bar{u}_t,\bar{X}_t) \\
=&\left\{
\begin{array}{ll}
-\zeta'_+ \big(\bar{u}_t^+  \big)
\varpi'_+\big(1-F_{\bar{u}_t^+}
(\bar{u}_t^+)\big)
&\mbox{ if }\bar{u}_t>0,\\
-\zeta'_-\big(\bar{u}_t^-\big)
\varpi'_-\big(1-F_{(\bar{u}^i_t)^-}
(\bar{u}_t^-)\big)
&\mbox{ if }\bar{u}_t<0,
 \end{array}
\right.
\end{align*}
a.e.$t\in[0,T]$,~a.s., where
\begin{equation*}
\left\{
\begin{array}{ccl}
dp_t&=&-\Big(b_x(t,\bar{u}_t,\bar{X}_t)p_t+\sigma_x(t,\bar{u}_t,\bar{X}_t) q_t
+\partial_x f(t,\bar{u}_t,\bar{X}_t)
\Big)dt+q_tdW_t,\\
p_T&=&l'(\bar{X}_T)w'(1-F_{\bar{X}_T}(\bar{X}_T)).
\end{array}
\right.
\end{equation*}
\end{remark}

We refer to Peng \cite{Peng 1990} for the classical optimal control problem of which the running cost function is $f(t,u_t,X_t)$.  The conclusion above can be proved by the same argument to be given in the next section, combining with those for classical stochastic maximum principle (see, e.g., Yong and Zhou \cite{Yong and Zhou 1999}).

\iffalse
\begin{align*}
J(u_\cdot)=&\mathbbm{E}\int^T_0 \Big(f(t,u_t,X_t)+
\sum^m_{i=1}\zeta^i_+({u^i_t}^+)
{\varpi^i_+}'\big(1-F_{{u^i_t}^+}({u^i_t}^+)\big) \\
&~~~-\sum^m_{i=1}\zeta^i_-({u^i_t}^-)
{\varpi^i_-}'\big(1-F_{{u^i_t}^-}({u^i_t}^-)\big)\Big)dt
+\mathbbm{E}\(l(X_T)w'\big(1-F_{X_T}(X_T)\big)\).
\end{align*}
\fi

\section{Proof of the Main Result}

In this section, we proceed to proving the stochastic maximum principle stated in Theorem \ref{thm}. The main idea is to perturb the optimal control in a careful way such that the sign of the control is not changed by the perturbation due to the singularity at 0. The key technique is in the study of the distribution functions of the perturbed state process evaluated at the state.

Suppose $\varepsilon\in[0,1)$. Take $u_\cdot\in\mathcal{U}$ such that $u_t$ has the same sign as $\bar u_t$ ($u_t=0$ if $\bar u_t=0$). Define
\[u^\varepsilon_\cdot=\bar{u}_\cdot+\varepsilon (u_\cdot-\bar{u}_\cdot).\]
 The convexity of $U$ guarantees that $u^\varepsilon_\cdot\in\mathcal{U}$, and obviously,
\[
J(\bar{u}_\cdot)-J(u^\varepsilon_\cdot)\ge 0.
\]
Denote the state trajectory corresponding to the perturbation $u^\varepsilon_\cdot$ of $\bar{u}_\cdot$ by $X^\varepsilon_\cdot$.

In the rest of this paper, we adopt the short-hand notations
\[v_t=u_t-\bar{u}_t,
~~\bar{\phi}(t)=\phi(t,\bar{u}_t,\bar{X}_t),\qquad \phi=b\mbox{ or }\si.\]

Now we proceed to proving Theorem $\ref{thm}$ by a few lemmas.
\begin{lemma} \label{lemma1}
Under Condition (H.1), we have
\[
\lim_{\varepsilon\to 0}\mathbbm{E}\big(\sup_{0\le t\le T} \big| X^\varepsilon_t-\bar{X}_t \big| ^4\big)=0.
\]
\end{lemma}
\textbf{Proof:}
From the state equation, one has
\begin{align*}
d\big(X^\varepsilon_t-\bar{X}_t\big)=&\Big(b(t,\bar{u}_t+\varepsilon v_t,X_t^\varepsilon)-b(t,\bar{u}_t,\bar{X}_t)\Big)dt\\
&+\Big(\sigma(t,\bar{u}_t+\varepsilon v_t,X_t^\varepsilon)-\sigma(t,\bar{u}_t,\bar{X}_t)\Big)dW_t.
\end{align*}
By Condition (H.1), Cauchy-Schwarz inequality as well as Burkholder-Davis-Gundy inequality, we obtain
\begin{align*}
\mathbbm{E}\sup_{0\le t\le T}\big|X^\varepsilon_t-\bar{X}_t\big|^4=&\mathbbm{E}\sup_{0\le t\le T}\Big\{\int^t_0\Big(b(s,\bar{u}_s+\varepsilon v_s,X_s^\varepsilon)-b(s,\bar{u}_s,\bar{X}_s)\Big)ds\\
&+\int^t_0\Big(\sigma(s,\bar{u}_s+\varepsilon v_s,X_s^\varepsilon)-\sigma(s,\bar{u}_s,\bar{X}_s)\Big)dW_s\Big\}^4\\
\le&8\mathbbm{E}\sup_{0\le t\le T}\Big\{\int^t_0\Big(b(s,\bar{u}_s+\varepsilon v_s,X_s^\varepsilon)-b(s,\bar{u}_s,\bar{X}_s)\Big)ds\Big\}^4\\
&+8\mathbbm{E}\sup_{0\le t\le T}\Big\{\int^t_0\Big(\sigma(s,\bar{u}_s+\varepsilon v_s,X_s^\varepsilon)-\sigma(s,\bar{u}_s,\bar{X}_s)\Big)dW_s\Big\}^4\\
\le&8\mathbbm{E}\sup_{0\le t\le T}\Big\{\int^t_0\Big(b(s,\bar{u}_s+\varepsilon v_s,X_s^\varepsilon)-b(s,\bar{u}_s,\bar{X}_s)\Big)^2ds\cdot\int^t_01^2ds\Big\}^2\\
&+8\(\frac43\)^4\mathbbm{E}\Big\{\int^T_0\Big(\sigma(s,\bar{u}_s+\varepsilon v_s,X_s^\varepsilon)-\sigma(s,\bar{u}_s,\bar{X}_s)\Big)^2ds\Big\}^2\\
\le&8T^3\mathbbm{E}\sup_{0\le t\le T}\int^t_0\Big(b(s,\bar{u}_s+\varepsilon v_s,X_s^\varepsilon)-b(s,\bar{u}_s,\bar{X}_s)\Big)^4ds\\
&+8T\(\frac43\)^4\mathbbm{E}\int^T_0\Big(\sigma(s,\bar{u}_s+\varepsilon v_s,X_s^\varepsilon)-\sigma(s,\bar{u}_s,\bar{X}_s)\Big)^4ds\\
\le&K_T\mathbbm{E}\int^T_0 (\big|X_s^\varepsilon-\bar{X}_s\big|+\varepsilon|v_s|)^4ds\\
\le& K_T\int^T_0\mathbbm{E}\sup_{0\le s\le t}\big|X_s^\varepsilon-\bar{X}_s\big|^4dt+K_T\mathbbm{E}\int^T_0\varepsilon^4|v_s|^4ds
\end{align*}
The result follows from Gronwall's inequality.
\qed

The following calculus lemma is a slight modification of Dini's theorem to suit our propose. We include it and its proof here for the completeness of this paper.

\begin{lemma}\label{lem0808a}
Suppose that $F_n$, $F$ are distibution functions on $\RR$ and $F$ is continuous. If for any $x$, $\lim_{n\to\infty}F_n(x)=F(x)$, then
\begin{equation}\label{eq0808a}
\lim_{n\to\infty}\sup_{x\in\RR}|F_n(x)-F(x)|=0.
\end{equation}
\end{lemma}
{\bf Proof:} If (\ref{eq0808a}) does not hold, then there exists $\ep_0>0$ and a sequence $x_n\in\RR$ such that $|F_n(x_n)-F(x)|\ge \ep_0$. Taking a subsequence if necessary, we may assume that $x_n\to x\in[-\infty,\infty]$ as $n\to\infty$. 

Suppose $x$ is finite. Taking subsequence if necessary, we may assume that $x_n\le x$ for all $n$, or $x_n\ge x$ for all $n$. We assume the former since the other case can be studied similarly. Further, taking subsequence if necessary, we may assume $x_n\uparrow x$. For $n\ge m$ large enough, we have
\[F_n(x_m)\le F_n(x_n)<F(x)-\ep_0.\]
Taking $n\to\infty$, we get $F(x_m)\le F(x)-\ep_0$. Taking $m\to\infty$, we then get $F(x)\le F(x)-\ep_0$ which is a contradiction.

Finally, we assume $x=\infty$ or $x=-\infty$. We take the former since the other is similar. Taking subsequence if necessary, we may assume that $x_n\uparrow\infty$. Let $n\ge m$ be large enough. Then,
\[F_n(x_m)\le F_n(x_n)<1-\ep_0.\]
Taking $n\to\infty$, we get $F(x_m)\le 1-\ep_0$. Letting $m\to\infty$, we arrive at the contradiction that $1\le 1-\ep_0$.

Since both cases lead to contradictions, (\ref{eq0808a}) must hold.
\qed

The following is the main technique lemma of this paper.

\begin{lemma}\label{lemma2}
Suppose that $\bar X_T$ possess continuous distribution function. Then,
\[
\lim_{\varepsilon\to 0}\mathbbm{E}\big|F_{X^\varepsilon_T}(X^\varepsilon_T)-F_{\bar{X}_T}(\bar{X}_T)\big|^4=0.
\]
\end{lemma}
\textbf{Proof:}
Notice that there is a subsequence $\varepsilon^*\subset\varepsilon$ such that
\[
\lim_{\varepsilon^*\to 0}\mathbbm{E}\big|F_{X^{\varepsilon^*}_T}(X^{\varepsilon^*}_T)-F_{\bar{X}_T}(\bar{X}_T)\big|^4
=\varlimsup_{\varepsilon\to 0}\mathbbm{E}\big|F_{X^\varepsilon_T}(X^\varepsilon_T)-F_{\bar{X}_T}(\bar{X}_T)\big|^4,
\]
which always exists. $X^\varepsilon_T \overset{L^4}{\to}\bar{X}_T$ implies that $X^{\varepsilon^*}_T \overset{L^4}{\to}\bar{X}_T$. Moreover, there is a subsequence 
$\bar{X}^{\varepsilon'}_t$ of $X^{\varepsilon^*}_T$ which converges to $\bar{X}_t$ almost surely. We then have
\[
\lim_{\varepsilon'\to 0}\mathbbm{E}\big|F_{X^{\varepsilon'}_T}(X^{\varepsilon'}_T)-F_{\bar{X}_T}(\bar{X}_T)\big|^4
=\lim_{\varepsilon^*\to 0}\mathbbm{E}\big|F_{X^{\varepsilon^*}_T}(X^{\varepsilon^*}_T)-F_{\bar{X}_T}(\bar{X}_T)\big|^4.
\]
As a consequence, the problem is turned to demonstrate
\begin{equation}\label{ep'}
\lim_{\varepsilon'\to 0}\mathbbm{E}\big|F_{X^{\varepsilon'}_T}(X^{\varepsilon'}_T)-F_{\bar{X}_T}(\bar{X}_T)\big|^4=0,
\end{equation}
given $X^{\varepsilon'}_T \overset{a.s.}{\to}\bar{X}_T$.
Note that
\begin{eqnarray*}
\big|F_{X^{\varepsilon'}_T}(X^{\varepsilon'}_T)-F_{\bar{X}_T}(\bar{X}_T)\big|
&\le& \big|F_{X^{\varepsilon'}_T}(X^{\varepsilon'}_T)
-F_{\bar{X}_T}(X^{\varepsilon'}_T)\big|
+\big|F_{\bar{X}_T}(X^{\varepsilon'}_T)
-F_{\bar{X}_T}(\bar{X}_T)\big|\\
&\le&\sup_x\left|F_{X^{\varepsilon'}_T}(x)-F_{\bar{X}_T}(x)\right|
+\big|F_{\bar{X}_T}(X^{\varepsilon'}_T)
-F_{\bar{X}_T}(\bar{X}_T)\big|\\
&\to&0,
\end{eqnarray*}
where the last step follows from Lemma \ref{lem0808a}, and the continuity of the distribution function $F_{\bar{X}_T}$. Equality $(\ref{ep'})$ then follows by the dominated convergence theorem.
\qed

\begin{remark}\label{rem0808a}
 One can also verify that for all $\lambda,\mu\in[0,1]$,
\begin{align*}
&\lim_{\varepsilon\to 0}\mathbbm{E} \big|\lambda F_{X^\varepsilon_T}(\mu X^\varepsilon_T+(1-\mu)\bar{X}_T)
 +(1-\lambda) F_{\bar{X}_T}(\mu X^\varepsilon_T+(1-\mu)\bar{X}_T)
 -F_{\bar{X}_T}(\bar{X}_T)\big|^4=0.
\end{align*}
Moreover,  if $\bar u^\pm_t$ have continuous (except at 0) distribution functions, then
\[
\lim_{\varepsilon\to 0}\mathbbm{E}\big|F_{(u^\varepsilon_t)^\pm}\big((u^\varepsilon_t)^\pm \big)-F_{\bar{u}_t^\pm}\big(\bar{u}_t^\pm\big)\big|^4=0,~~\forall t\in[0,T].
\]
\end{remark}

The next lemma provides the first order perturbation of the state process.

\begin{lemma}\label{EZ}
Let $Z_t$ be such that
\begin{equation}\label{eqz}
\left\{
\begin{array}{ccl}
 dZ_t & =  &( \bar{b}_x(t)Z_t+\bar{b}_u(t) v_t)dt
+ (\bar{\sigma}_x(t) Z_t+\bar{\sigma}_u(t)v_t)dW_t  \\
  Z_0 & =  & 0.
\end{array}
\right.
\end{equation}
Then, under Condition (H.1), we have
\begin{equation}
\lim_{\varepsilon\to 0}\mathbbm{E}\Big(\sup_{0\le t\le T} \Big| \frac{X^\varepsilon_t-\bar{X}_t}{\varepsilon}-Z_t \Big| ^2\Big)=0.
\end{equation}
\end{lemma}
\textbf{Proof:}
Let $y^\varepsilon_t= \frac{X^\varepsilon_t-\bar{X}_t}{\varepsilon}-Z_t$, then
\begin{align*}
dy^\varepsilon_t=&\Big\{\frac{1}{\varepsilon}\Big(b(t,\bar{u}_t+\varepsilon v_t,\bar{X}_t+\varepsilon(Z_t+y^\varepsilon_t))-b(t,\bar{u}_t,\bar{X}_t)\Big)-b_x(t,\bar{u}_t,\bar{X}_t)Z_t\\
&-b_u(t,\bar{u}_t,\bar{X}_t)v_t\Big\}dt
+\Big\{\frac{1}{\varepsilon}\Big(\sigma(t,\bar{u}_t+\varepsilon v_t,\bar{X}_t+\varepsilon(Z_t+y^\varepsilon_t))-\sigma(t,\bar{u}_t,\bar{X}_t)\Big)\\
&-\sigma_x(t,\bar{u}_t,\bar{X}_t)Z_t-\sigma_u(t,\bar{u}_t,\bar{X}_t)v_t\Big\}dW_t.
\end{align*}
One can easily show that $\mathbbm{E}\int^T_0 Z_t^4dt+\mathbbm{E}\int^T_0 |y_t^\varepsilon|^4dt<\infty$. Since the drift and the diffusion coefficients of $y^\varepsilon_t$ are similar, we focus on the drift one only. Note that
\begin{align*}
&\frac{1}{\varepsilon}\Big(b(t,\bar{u}_t+\varepsilon v_t,\bar{X}_t+\varepsilon(Z_t+y^\varepsilon_t))-b(t,\bar{u}_t,\bar{X}_t)\Big)
-b_x(t,\bar{u}_t,\bar{X}_t)Z_t-b_u(t,\bar{u}_t,\bar{X}_t)v_t\\
&=\int^1_0 b_x(t,\bar{u}_t+\lambda\varepsilon v_t,\bar{X}_t+\lambda\varepsilon(Z_t+y^\varepsilon_t))(Z_t+y^\varepsilon_t)d\lambda\\
&~~~~~+\int^1_0 b_u(t,\bar{u}_t+\lambda\varepsilon v_t,\bar{X}_t+\lambda\varepsilon(Z_t+y^\varepsilon_t))v_td\lambda
-b_x(t,\bar{u}_t,\bar{X}_t)Z_t-b_u(t,\bar{u}_t,\bar{X}_t)v_t\\
&=\int^1_0 \Big(b_x(t,\bar{u}_t+\lambda\varepsilon v_t,\bar{X}_t+\lambda\varepsilon(Z_t+y^\varepsilon_t))
-b_x(t,\bar{u}_t,\bar{X}_t)\Big)Z_td\lambda\\
&~~~~~+\int^1_0 \Big(b_u(t,\bar{u}_t+\lambda\varepsilon v_t,\bar{X}_t+\lambda\varepsilon(Z_t+y^\varepsilon_t))
-b_u(t,\bar{u}_t,\bar{X}_t)\Big)v_td\lambda\\
&~~~~~+\int^1_0 b_x(t,\bar{u}_t+\lambda\varepsilon v_t,\bar{X}_t+\lambda\varepsilon(Z_t+y^\varepsilon_t))y^\varepsilon_td\lambda.
\end{align*}
By using Condition (H.1) as well as Cauchy-Schwarz inequality, we conclude that the first two terms on the right hand side of the above equality tend to zero in $L^2(\Omega\times[0,T])$ as $\varepsilon$ goes to zero. In fact, the first term is estimated as follows:
\begin{align*}
&\mathbbm{E}\int^T_0\Big\{\int^1_0 \Big(b_x(t,\bar{u}_t+\lambda\varepsilon v_t,\bar{X}_t+\lambda\varepsilon(Z_t+y^\varepsilon_t))
-b_x(t,\bar{u}_t,\bar{X}_t)\Big)Z_td\lambda\Big\}^2dt\\
&\le\mathbbm{E}\int^T_0\Big\{\int^1_0 K\lambda\varepsilon\big(|Z_t+y^\varepsilon_t|+|v_t|\big)Z_td\lambda\Big\}^2dt\\
&\le\mathbbm{E}\int^T_0K\Big\{\int^1_0 \lambda\varepsilon\big(|Z_t+y^\varepsilon_t|+|v_t|\big)d\lambda\Big\}^2\cdot Z_t^2dt\\
&\le\mathbbm{E}\int^T_0K\int^1_0 \big\{\lambda\varepsilon\big(|Z_t+y^\varepsilon_t|+|v_t|\big)\big\}^2d\lambda\cdot\int^1_0 1^2d\lambda\cdot Z_t^2dt\\
&\le\mathbbm{E}\int^T_0\Big\{K\int^1_0 \big(\lambda\varepsilon|Z_t+y^\varepsilon_t|\big)^2d\lambda
+K\int^1_0 (\lambda\varepsilon v_t)^2d\lambda \Big\}\cdot Z_t^2dt\\
&\le K\int^T_0\mathbbm{E}\int^1_0 \big(\lambda\varepsilon|Z_t+y^\varepsilon_t|\big)^2d\lambda\cdot Z_t^2dt
+K\int^T_0\mathbbm{E}\int^1_0 (\lambda\varepsilon v_t)^2d\lambda\cdot Z_t^2dt\\
&\le K\int^T_0\Big\{\mathbbm{E}\Big[\int^1_0 \big(\lambda\varepsilon|Z_t+y^\varepsilon_t|\big)^2d\lambda\Big]^2
\cdot\mathbbm{E} Z_t^4\Big\}^{\frac{1}{2}}dt\\
&~~~~+K\Big\{\int^T_0\mathbbm{E}\Big[\int^1_0 (\lambda\varepsilon v_t)^2d\lambda\Big]^2 dt
\Big\}^{\frac{1}{2}}
\cdot\Big\{\int^T_0 \mathbbm{E} Z_t^4dt\Big\}^{\frac{1}{2}}\\
&\le K\Big\{\int^T_0\mathbbm{E}\int^1_0 \big(\lambda\varepsilon|Z_t+y^\varepsilon_t|\big)^4d\lambda dt
\Big\}^{\frac{1}{2}}
\cdot\Big\{\int^T_0\mathbbm{E} Z_t^4dt\Big\}^{\frac{1}{2}}\\
&~~~~+K\Big\{\int^T_0\mathbbm{E} \int^1_0 (\lambda\varepsilon v_t)^4d\lambda dt
\Big\}^{\frac{1}{2}}
\cdot\Big\{\int^T_0\mathbbm{E} Z_t^4dt\Big\}^{\frac{1}{2}}\to 0 \mbox{ as }\varepsilon\to 0.
\end{align*}
The proof for the second term is similar. Dealing with the diffusion part of $y^\varepsilon_t$ by the same treatment, one has
\begin{align*}
y^\varepsilon_t
=&\int^t_0\int^1_0 b_x(s,\bar{u}_s+\lambda\varepsilon v_s,\bar{X}_s+\lambda\varepsilon(Z_s+y^\varepsilon_s))y^\varepsilon_sd\lambda ds+\int^t_0\rho_s^\varepsilon ds\\
&+\int^t_0\int^1_0 \sigma_x(s,\bar{u}_s+\lambda\varepsilon v_s,\bar{X}_s+\lambda\varepsilon(Z_s+y^\varepsilon_s))y^\varepsilon_sd\lambda dW_s+\int^t_0\tau_s^\varepsilon dW_s,
\end{align*}
where $\mathbbm{E}\int^T_0|\rho^\varepsilon_t|^2 dt,~\mathbbm{E}\int^T_0|\tau^\varepsilon_t|^2 dt$ go to zero as $\varepsilon$ goes to zero.  Using the Burkholder-Davis-Gundy inequality, in addition to the boundedness condition of $b_x, \sigma_x$, finally we have
\begin{align*}
\mathbbm{E}\sup_{0\le t\le T}|y^\varepsilon_t|^2
\le&K\int^T_0\sup_{0\le s\le t}|y^\varepsilon_s|^2 dt+\mathbbm{E}\int^T_0|\rho_s^\varepsilon|^2ds+\mathbbm{E}\int^T_0|\tau_s^\varepsilon|^2ds.
\end{align*}
Applying Gronwall's inequality, the result then follows.
\qed

In the next lemma, we calculate the derivative of the perturbed prospective functional with respect to $\varepsilon$.

\begin{lemma}\label{lemmaj1}
The Gateaux derivative of the objective functional $J$ is given by
\begin{align*}
\frac{d}{d\varepsilon}J(\bar{u}_\cdot+\varepsilon v_\cdot)\big|_{\varepsilon=0}
&=\mathbbm{E}\int^T_0\Big(\zeta'_+\big(\bar{u}^+_t\big)
\varpi'_+\big(1-F_{\bar{u}^+_t}(\bar{u}^+_t)\big)v_t\mathbbm{1}_{\bar{u}_t>0}\\
&~~~~~~~~~~~~~~+\zeta'_-\big(\bar{u}^-_t\big)
\varpi'_-\big(1-F_{\bar{u}^-_t}(\bar{u}^-_t)\big)v_t\mathbbm{1}_{\bar{u}_t<0}\\
&~~~+\mathbbm{E}\Big(l'(\bar{X}_T)w'\big(1-F_{\bar{X}_T}(\bar{X}_T)\big)Z_T\Big).
\end{align*}\end{lemma}
\textbf{Proof:} Recalling $(\ref{eqJ0})$, the
 three integrals of the objective functional are similar in structure. We discuss the last term in details. Rewrite that term as
\begin{align*}
\mathbbm{E} \big(l(X_T)w'\big(1-F_{X_T}(X_T)\big)\big)
&=\int_0^{\infty}l(x)w'\big(1-F_{X_T}(x)\big)dF_{X_T}(x)\\
&=\int_0^{\infty}l'(x)w(1-F_{X_T}(x))dx.
\end{align*}
We now calculate its Gateaux derivative as
\begin{align*}
I\overset{\triangle}{=}&\lim_{\varepsilon\to 0}\frac{1}{\varepsilon}\Big(\int_0^{\infty}l'(x)w(1-F_{X_T^\varepsilon}(x))dx-
\int_0^{\infty}l'(x)w(1-F_{\bar{X}_T}(x))dx\Big)\\
=&\lim_{\varepsilon\to 0}\frac{1}{\varepsilon}\int_0^{\infty}l'(x)\int^1_0w'(1-\lambda F_{X_T^\varepsilon}(x)
-(1-\lambda)F_{\bar{X}_T}(x))d\lambda(F_{\bar{X}_T}(x)-F_{X_T^\varepsilon}(x)) dx,
\end{align*}
For convenience, let
\[
g_\varepsilon(x)=
l'(x)\int^1_0w'_+(1-\lambda F_{X_T^\varepsilon}(x)
-(1-\lambda)F_{\bar{X}_T}(x))d\lambda,~x>0,
\]
and define
\[
G_\varepsilon(x)=\int^x_0 g_\varepsilon(y)dy, ~G_\varepsilon(0+)=0.
\]
Then,
\begin{align*}
I=&\lim_{\varepsilon\to 0}\varepsilon^{-1}\int_{0}^{\infty}g_\varepsilon(x)(F_{\bar{X}_T}(x)-F_{X_T^\varepsilon}(x))dx\\
=&\lim_{\varepsilon\to 0}\varepsilon^{-1}\int_{0}^{\infty}(F_{\bar{X}_T}(x)-F_{X_T^\varepsilon}(x))dG_\varepsilon(x)\\
=&-\lim_{\varepsilon\to 0}\varepsilon^{-1}\int_{0}^{\infty}G_\varepsilon(x)d(F_{\bar{X}_T}(x)-F_{X_T^\varepsilon}(x))\\
=&\lim_{\varepsilon\to 0}\varepsilon^{-1}\mathbbm{E}(G_\varepsilon(X_T^\varepsilon)-G_\varepsilon(\bar{X}_T))\\
=&\lim_{\varepsilon\to 0}\varepsilon^{-1}\mathbbm{E}\int^1_0g_\varepsilon(\mu X_T^\varepsilon+(1-\mu)\bar{X}_T)(X_T^\varepsilon-\bar{X}_T)d\mu.
\end{align*}

The other terms can be studied similarly. Hence the Gateaux derivative of $J$ is translated to be
\begin{align}\label{*}
\frac{d}{d\varepsilon}J(\bar{u}_\cdot+\varepsilon v_\cdot)\big|_{\varepsilon=0}
=&\lim_{\varepsilon\to 0}\frac{1}{\ep}\mathbbm{E}\int^T_0\int^1_0
g^1_\varepsilon(\tau u^\ep_t +(1-\tau)\bar{u}_t)
(u^\ep_t-{\bar{u}_t})
\mathbbm{1}_{\bar{u}_t>0}d\tau dt \nonumber\\
&+\lim_{\varepsilon\to 0}\frac{1}{\ep}\mathbbm{E}\int^T_0\int^1_0g^2_\varepsilon
(-\tau u^\ep_t-(1-\tau)\bar{u}_t)
(u^\ep_t-{\bar{u}_t})
\mathbbm{1}_{\bar{u}_t<0}d\tau dt \nonumber \\
&+\lim_{\varepsilon\to 0}\frac{1}{\ep}\mathbbm{E}\int^1_0g_\varepsilon(\mu X_T^\varepsilon+(1-\mu)\bar{X}_T)(X_T^\varepsilon-\bar{X}_T)d\mu,
\end{align}
where
\[
g^1_\varepsilon(x)=
\zeta'_+(x)\int^1_0\varpi'_+(1-\lambda F_{(u^\ep_t)^+}(x)
-(1-\lambda)F_{\bar{u}_t^+}(x))d\lambda,~x>0,
\]
\[
g^2_\varepsilon(x)=
\zeta'_-(x)\int^1_0\varpi'_-(1-\lambda F_{(u^\ep_t)^-}(x)
-(1-\lambda)F_{\bar{u}_t^-}(x))d\lambda,~x>0.
\]

%--------------------------
\begin{comment}
We discover that
\begin{align*}
&\lim_{\varepsilon\to 0}\int^1_0g_\varepsilon\(\mu X_T^\varepsilon+(1-\mu)\bar{X}_T\)d\mu\\
&=\lim_{\varepsilon\to 0}\int^1_0l'\(\mu X_T^\varepsilon+(1-\mu)\bar{X}_T\)
\cdot\int^1_0w'\Big(1-\lambda F_{\widetilde{X_T^\varepsilon}}\(\mu X_T^\varepsilon+(1-\mu)\bar{X}_T\)\\
&~~~   -(1-\lambda)F_{\widetilde{\bar{X}_T}}\(\mu X_T^\varepsilon+(1-\mu)\bar{X}_T\)\Big)d\lambda d\mu\\
&=\int^1_0l'(\bar{X}_T)\int^1_0w'\Big(1-
\lambda F_{\widetilde{\bar{X}_T}}(\bar{X}_T)
-(1-\lambda)F_{\widetilde{\bar{X}_T}}(\bar{X}_T)\Big)d\lambda d\mu\\
&=l'(\bar{X}_T)w'\Big(1-F_{\widetilde{\bar{X}_T}}(\bar{X}_T)\Big).
\end{align*}
\end{comment}
%-------------------------
Next, we go back to the calculation of $I$. To prove
\[
\lim_{\varepsilon\to 0}\mathbbm{E}\Big|\int^1_0g_\varepsilon\(\mu X_T^\varepsilon+(1-\mu)\bar{X}_T\)d\mu-
l'(\bar{X}_T)w'\Big(1-F_{\bar{X}_T}(\bar{X}_T)\Big)\Big|^2= 0,
\]
we adopt the shorthand notation for simplicity,
\[J^{\varepsilon,\lambda,\mu}=\lambda F_{X_T^\varepsilon}\big(\mu X_T^\varepsilon+(1-\mu)\bar{X}_T\big)
 +(1-\lambda)F_{\bar{X}_T}\big(\mu X_T^\varepsilon+(1-\mu)\bar{X}_T\big),
 \]
\[
J_1=w'\big(1-J^{\varepsilon,\lambda,\mu}\big)-w'\big(1-F_{\bar{X}_T}(\bar{X}_T)\big).
\]
Condition (H.3) implies that $J_1$ is bounded. Therefore, by Cauchy-Schwarz inequality,
\begin{align*}
&\lim_{\varepsilon\to 0}\mathbbm{E}\Big|\int^1_0g_\varepsilon\(\mu X_T^\varepsilon+(1-\mu)\bar{X}_T\)d\mu
-l'(\bar{X}_T)w'\Big(1-F_{\bar{X}_T}(\bar{X}_T)\Big)\Big|^2\\
&\le\lim_{\varepsilon\to 0}\mathbbm{E}\Big|\int^1_0 l'\big(\mu X_T^\varepsilon+(1-\mu)\bar{X}_T\big)\int^1_0
w'\big(1-\lambda F_{X_T^\varepsilon}\big(\mu X_T^\varepsilon+(1-\mu)\bar{X}_T\big)\\
&~~~~-(1-\lambda)F_{\bar{X}_T}\big(\mu X_T^\varepsilon+(1-\mu)\bar{X}_T\big)\big)d\lambda d\mu
-l'(\bar{X}_T)w'\big(1-F_{\bar{X}_T}(\bar{X}_T)\big)\Big|^2\\
&\le\lim_{\varepsilon\to 0} K\mathbbm{E}\Big|\int^1_0 \Big(l'\big(\mu X_T^\varepsilon+(1-\mu)\bar{X}_T\big)-l'(\bar{X}_T)\Big)
\cdot\int^1_0w'\big(1-J^{\varepsilon,\lambda,\mu}\big)d\lambda d\mu\Big|^2\\
&~~~~  +\lim_{\varepsilon\to 0} K\mathbbm{E} \big| l'(\bar{X}_T)\big|^2\cdot
\Big|\int^1_0 \int^1_0
w'\big(1-J^{\varepsilon,\lambda,\mu}\big)d\lambda d\mu
-w'\big(1-F_{\bar{X}_T}(\bar{X}_T)\big)\Big|^2\\
&\le\lim_{\varepsilon\to 0} K\mathbbm{E}\int^1_0 \int^1_0l''\big(\tau\mu X_T^\varepsilon+(1-\tau\mu)\bar{X}_T\big)^2
(X_T^\varepsilon-\bar{X}_T)^2\mu^2 d\tau d\mu\\
&~~~~ +K\Big(\mathbbm{E} \big| l'(\bar{X}_T)\big|^4\Big)^{\frac{1}{2}}\cdot
\lim_{\varepsilon\to 0} \Big(\mathbbm{E}\int^1_0 \int^1_0\Big|
w'\big(1-J^{\varepsilon,\lambda,\mu}\big)
-w'\big(1-F_{\bar{X}_T}(\bar{X}_T)\big)\Big|^4d\lambda d\mu \Big)^{\frac{1}{2}}.
\end{align*}
Thanks to Lemma $\ref{lemma1}$, we obtain
\begin{align*}
&\lim_{\varepsilon\to 0} K\mathbbm{E}\int^1_0 \int^1_0l''\big(\tau\mu X_T^\varepsilon+(1-\tau\mu)\bar{X}_T\big)^2
(X_T^\varepsilon-\bar{X}_T)^2\mu^2 d\tau d\mu\\
&\le \lim_{\varepsilon\to 0} K\mathbbm{E}\big(l''(X_T^\varepsilon)+l''(\bar{X}_T)\big)^2
(X_T^\varepsilon-\bar{X}_T)^2\\
&\le \lim_{\varepsilon\to 0} K\big(\mathbbm{E}\big(l''(X_T^\varepsilon)+l''(\bar{X}_T)\big)^4
\mathbbm{E}(X_T^\varepsilon-\bar{X}_T)^4\big)^{\frac{1}{2}}=0.
\end{align*}
Meanwhile, acccording to the Remark \ref{rem0808a}, we have
\begin{eqnarray*}
I_2&\overset{\triangle}{=}&\lim_{\varepsilon\to 0} \mathbbm{E}\int^1_0 \int^1_0\Big|
w'\big(1-J^{\varepsilon,\lambda,\mu}\big)
-w'\big(1-F_{\bar{X}_T}(\bar{X}_T)\big)\Big|^4d\lambda d\mu \\
&\le& \lim_{\varepsilon\to 0} \mathbbm{E}\int^1_0 \int^1_0\Big|\int^1_0
w''\Big(1-\tau J^{\varepsilon,\lambda,\mu}-(1-\tau)F_{\bar{X}_T}(\bar{X}_T)\Big)d\tau
\big(F_{\bar{X}_T}(\bar{X}_T)-J^{\varepsilon,\lambda,\mu}\big)\Big|^4\\
&&~~~~~~~~~~~~~~~~~~~
~~~~~~~~~\cdot\mathbbm{1}_{\delta\le J^{\varepsilon,\lambda,\mu},F_{\bar{X}_T}(\bar{X}_T)\le 1-\delta}
d\lambda d\mu\\
&&+\lim_{\varepsilon\to 0} \mathbbm{E}\int^1_0 \int^1_0 | J_1 |^4
\cdot\mathbbm{1}_{J^{\varepsilon,\lambda,\mu}<\delta} d\lambda d\mu
+\lim_{\varepsilon\to 0} \mathbbm{E}\int^1_0 \int^1_0 | J_1 |^4
\cdot\mathbbm{1}_{F_{\bar{X}_T}(\bar{X}_T)<\delta} d\lambda d\mu\\
&&+\lim_{\varepsilon\to 0} \mathbbm{E}\int^1_0 \int^1_0 | J_1 |^4
\cdot\mathbbm{1}_{J^{\varepsilon,\lambda,\mu}>1-\delta} d\lambda d\mu
+\lim_{\varepsilon\to 0} \mathbbm{E}\int^1_0 \int^1_0 | J_1 |^4
\cdot\mathbbm{1}_{F_{\bar{X}_T}(\bar{X}_T)>1-\delta} d\lambda d\mu\\
&\le& \lim_{\varepsilon\to 0}K_{\delta} \mathbbm{E}\int^1_0 \int^1_0
\big|F_{\bar{X}_T}(\bar{X}_T)-J^{\varepsilon,\lambda,\mu}\big|^4
d\lambda d\mu
+\lim_{\varepsilon\to 0} K\mathbbm{E}\int^1_0 \int^1_0
\mathbbm{1}_{J^{\varepsilon,\lambda,\mu}<\delta} d\lambda d\mu\\
&&+\lim_{\varepsilon\to 0} K\mathbbm{E}\int^1_0 \int^1_0
\mathbbm{1}_{J^{\varepsilon,\lambda,\mu}>1-\delta} d\lambda d\mu
+K\mathbbm{E}( \mathbbm{1}_{F_{\bar{X}_T}(\bar{X}_T)<\delta})
+ K\mathbbm{E} (\mathbbm{1}_{F_{\bar{X}_T}(\bar{X}_T)>1-\delta} )\\
&=& K\int^1_0 \int^1_0
\lim_{\varepsilon\to 0}P\big\{J^{\varepsilon,\lambda,\mu}<\delta\big\} d\lambda d\mu
+ K P\{ F_{\bar{X}_T}(\bar{X}_T)<\delta\}\\
&&+ K\int^1_0 \int^1_0
\lim_{\varepsilon\to 0}P\big\{J^{\varepsilon,\lambda,\mu}>1-\delta\big\} d\lambda d\mu
+ K P\{F_{\bar{X}_T}(\bar{X}_T)>1-\delta \}\\
&\le& 2K P\{F_{\bar{X}_T}(\bar{X}_T)\le \delta)\} +2KP\{F_{\bar{X}_T}(\bar{X}_T)\ge 1-\delta\}.
\end{eqnarray*}
It is recognized that the random variable $F_{\bar{X}_T}(\bar{X}_T)\sim U(0,1)$. Taking $\delta\to 0$, we see that $I_2$ is 0. Consequently,
\[
\lim_{\varepsilon\to 0}\mathbbm{E}\Big|\int^1_0g_\varepsilon\(\mu X_T^\varepsilon+(1-\mu)\bar{X}_T\)d\mu-
l'(\bar{X}_T)w'\Big(1-F_{\bar{X}_T}(\bar{X}_T)\Big)\Big|^2= 0.
\]
By Lemma $\ref{EZ}$, we then arrive at
\[
I=\mathbbm{E}\Big(l'(\bar{X}_T)w'(1-F_{\bar{X}_T}(\bar{X}_T))Z_T\Big).
\]

Other terms on the RHS of $(\ref{*})$ can be treated by the same way.
\qed

As the last technique step, we write $I$ above into a form which is the same as the other two terms in the derivative of the prospective functional given in last lemma.

\begin{lemma}\label{lemmapz}
\[
\mathbbm{E}(p_TZ_T)=\mathbbm{E}\int^T_0v_t(p_t\bar{b}_u(t)+q_t\bar{\sigma}_u(t))dt.
\]
\end{lemma}
$\textbf{Proof:}$
In view of $(\ref{eqz})$ and $(\ref{adjoint})$, applying It\^o's formula to $p_tZ_t$ yeilds
\begin{align*}
d(p_tZ_t)=&p_tdZ_t+Z_tdp_t+d\left<p,Z\right>_t\\
=&\big(p_t\bar{b}_x(t)Z_t+p_t\bar{b}_u(t) v_t\big)dt+p_t\big(\bar{\sigma}_x(t) Z_t+\bar{\sigma}_u(t)v_t\big)dW_t\\
&-Z_t\big(\bar{b}_x(t)p_t+\bar{\sigma}_x(t) q_t\big)dt
+Z_tq_tdW_t
+\big(\bar{\sigma}_x(t) Z_t+\bar{\sigma}_u(t)v_t\big)q_tdt\\
=&\big(p_t\bar{b}_u(t) v_t+\bar{\sigma}_u(t)v_tq_t \big)dt
+\big(p_t\bar{\sigma}_x(t) Z_t+p_t\bar{\sigma}_u(t)v_t+Z_tq_t\big)dW_t.
\end{align*}
Then, taking the integration over $t$ and taking the expectation on both side, the result follows.
\qed

Finally, we are ready to finish 

{\bf Proof of Theorem \ref{thm}}: Combining Lemma $\ref{lemmaj1}$ and $\ref{lemmapz}$,
the Gateaux derivative of the prospective functional is expressed in this way.
\begin{align*}
&\frac{d}{d\varepsilon}J(\bar{u}_\cdot+\varepsilon v_\cdot)\Big|_{\varepsilon=0}
=\mathbbm{E}\int^T_0
v_t\Big(p_t\bar{b}_u(t)+\bar{\sigma}_u(t)q_t
+\zeta'_+\big(\bar{u}^+_t\big)
\varpi'_+\big(1-F_{\bar{u}^+_t}(\bar{u}^+_t)\big)\mathbbm{1}_{\bar{u}_t>0}\\
&~~~~~~~~~~~~~~~~~~~~~~~~~~~~~~~~~~
+\zeta'_-\big(\bar{u}^-_t\big)
\varpi'_-\big(1-F_{\bar{u}^-_t}(\bar{u}^-_t)\big)\mathbbm{1}_{\bar{u}_t<0}\Big)dt.
\end{align*}
Since $\bar{u}_\cdot$ is optimal, we arrive at
\begin{align*}
\mathbbm{E}\int^T_0
&(u_t-\bar{u}_t)\Big(p_t\bar{b}_u(t)+\bar{\sigma}_u(t)q_t
+\zeta'_+\big(\bar{u}^+_t\big)
\varpi'_+\big(1-F_{\bar{u}^+_t}(\bar{u}^+_t)\big)\mathbbm{1}_{\bar{u}_t>0}\\
&~~~~~~~~~~~~~~~~~~~~~~~~~~~~
+\zeta'_-\big(\bar{u}^-_t\big)
\varpi'_-\big(1-F_{\bar{u}^-_t}(\bar{u}^-_t)\big)\mathbbm{1}_{\bar{u}_t<0}\Big)dt=0.
\end{align*}
Note that, when $\bar u_t\neq 0$,  $u_t-\bar u_t$ is arbitrary. Thus, in this case, we have
\begin{align*}
&p_t\bar{b}_u(t)+\bar{\sigma}_u(t)q_t
+\zeta'_+\big(\bar{u}^+_t\big)
\varpi'_+\big(1-F_{\bar{u}^+_t}
(\bar{u}^+_t)\big)\mathbbm{1}_{\bar{u}_t>0}\\
&~~~~~~~~~~~~~~~~~~~~~~~~~~~~~~
+\zeta'_-\big(\bar{u}^-_t\big)
\varpi'_-\big(1-F_{\bar{u}^-_t}(\bar{u}^-_t)\big)\mathbbm{1}_{\bar{u}_t<0}=0,
\end{align*}
a.e.$t\in[0,T]$, $\mathbbm{P}$-a.s..  
\qed

\section{Application}

In this section, we apply our maximum principle to three interesting examples. The first example will show that the result in Jin and Zhou \cite{Jin and Zhou 2008} coincides with ours when the running cost is absent. The other two examples demonstrate that some optimization problems can be solved explicitly using our stochastic maximum principle.

\begin{example}
Consider Example 1.1 without consumption. The state process is modeled by
\[
\left\{
\begin{array}{ccl}
dX_t&=&X_t(r_t+(b_t-r_t\mathbf{1_m})^\top_tu_t)dt+X_tu^\top_t\sigma_tdW_t,~~~t\in[0,T];\\
X_0&=&x_0>0,
\end{array}
\right.
\]
and the agent's objective functional under the CPT becomes
\[
V(X_T)=\int^{\infty}_0w(\mathbbm{P}\{l(X_T)>x\})dx.
\]
\textbf{Hypothesis.}  There exists an $\mathbbm{R}^m$-valued, uniformly bounded, $\mathcal{F}_t$-progressively measurable process $\theta.$ such that $\sigma_t\theta_t=b_t-r_t\mathbf{1_m}$, a.e.$t\in[0,T]$, a.s.. Besides, rank$(\sigma_t)=m$, a.e.$t\in[0,T]$, a.s..

The Hypothesis ensures that the financial market is arbitrage-free and complete. Under suitable conditions, the optimal terminal wealth given by Jin and Zhou \cite{Jin and Zhou 2008} (section 6) is
\begin{equation}\label{ZX}
\bar{X}_T=
(l')^{-1}\Big(\frac{\lambda\rho_T}{w'(F_{\rho_T}(\rho_T))}\Big),
\end{equation}
where
\[
\rho_t=\exp\Big\{-\int^t_0\big(r_s+\frac{1}{2}|\theta_s|^2\big)ds-\int^t_0\theta^\top_sdW_s\Big\}
\]
is the pricing kernel, and $\lambda>0$ is the unique real number such that $\mathbbm{E}(\rho_T\bar{X}_T)=x_0$.
And they proved
\[
F_{\rho_T}(\rho_T)=1-F_{\bar{X}_T}(\bar{X}_T).
\]

In the light of Theorem $\ref{thm}$, an optimal solution $(\bar{u}_\cdot,\bar{X}_\cdot)$ must satisfy $(\ref{adjoint})$ and $(\ref{smp})$. In fact, substituting $(\ref{ZX})$ into $(\ref{adjoint})$, we are able to obtain
\[
\left\{
\begin{array}{ccl}
dp_t&=&-(r_t+(b_t-r_t\mathbf{1_m})^\top_tu_t)p_tdt-u^\top_t\sigma_tq_tdt+q^\top_tdW_t,\\
p_T&=&l'(\bar{X}_T)w'(1-F_{\bar{X}_T}(\bar{X}_T))=\lambda\rho_T.
\end{array}
\right.
\]
Applying  It\^o's formula to $\lambda\rho_t$, one has
\[
d\big(\lambda\rho_t)
=-r_t\cdot\lambda\rho_tdt-\lambda\rho_t\theta^\top_tdW_t.
\]
Comparing it with the above backward SDE, it yields
\[
p_t=\lambda\rho_t,
~~~
q_t=-\lambda\rho_t\theta_t.
\]
With $\sigma_t\theta_t=b_t-r_t\mathbf{1_m}$, we achieve
\[
p_t (b_t-r_t\mathbf{1_m})+\sigma_tq_t=
\lambda\rho_t (b_t-r_t\mathbf{1_m})-\sigma_t\lambda\rho_t\theta_t=0,~\forall t\in[0,T],
\]
namely, $(p,q)$ satisfies $(\ref{smp})$. In other words, the optimal strategy obtained in this paper consists with that of Jin and Zhou \cite{Jin and Zhou 2008}.
\end{example}

Actually, some situation would lead to no solution when $\bar{u}_t\neq 0$. In other words, the unique solution to the control process is 0.
\begin{example}
Let $b(t,u,x)=-ux$, $\sigma(t,u,x)=x$. Suppose there is no terminal term in objective functional; namely,
\begin{equation*}
J(u_\cdot)=\mathbbm{E}\int^T_0 \Big(\zeta_+(u_t^+)\varpi'_+\big(1-F_{u^+_t}(u^+_t)\big)
-\zeta_-(u^-_t)\varpi'_-\big(1-F_{u^-_t}(u^-_t)\big)\Big)dt.
\end{equation*}

If $(\bar{u}_\cdot,\bar{X}_\cdot)$ is an optimal solution, by the stochastic maximum principle, we have
\begin{eqnarray*}
p_t=
\left\{
\begin{array}{ll}
\zeta'_+\big(\bar{u}^+_t\big)
\varpi'_+\big(1-F_{\bar{u}^+_t}
(\bar{u}^+_t)\big)
&\mbox{ if }\bar{u}_t>0,\\
\zeta'_-\big(\bar{u}^-_t\big)
\varpi'_-\big(1-F_{\bar{u}^-_t}
(\bar{u}^-_t)\big)
&\mbox{ if }\bar{u}_t<0,
 \end{array}
\right. a.s..
\end{eqnarray*}
On the other hand, $(p_t,q_t)$ solves the BSDE 
\begin{equation*}
\left\{
\begin{array}{ccl}
dp_t&=&\big(\bar{u}_t p_t- q_t
\big)dt+q_tdW_t,\\
p_T&=&0.
\end{array}
\right.
\end{equation*}
Clearly, $p_t\equiv q_t\equiv 0$ is the unique solution, which results in a contradiction if $\bar{u}_t\neq 0$. Accordingly,  $\bar{u}_t=0,~a.e.\;\ t\in[0,T], a.s.$.
\end{example}

Finally, we present a solvable example and compare the result with the one without probability distortions. The process $u_t^\pm$ in the objective functional are replaced by $u_t^\pm X_t$, signifying the proportion of wealth process.  We study a case with compounded cost function.
\begin{example}\label{ex3}
Let $u_t,X_t>0$, $b(t,u,x)=-ux$, $\sigma(t,u,x)=x$, $\zeta_+(x)=\frac{x^\alpha}{\alpha}(0<\alpha<1)$, $\varpi_+(p)=\nu p^{\gamma+1}+(1-\nu)[1-(1-p)^{\beta+1}](\gamma,\beta\ge 0,0\le\nu\le 1)$.
%$\varpi_-(p)=p^{\beta+1}(\beta\ge 0)$
We have
\begin{equation*}
\left\{
\begin{array}{ccl}
 dX_t  &=&   -u_tX_tdt+X_t dW_t,   \\
  X_0  &=&  x_0,
\end{array}
\right.
\end{equation*}
and
\begin{align*}
J(u_\cdot)=&\mathbbm{E}\int^T_0
\Big(\frac{1}{\alpha}(u_tX_t)^\alpha\varpi_+'\big(1-F_{u_tX_t}(u_tX_t)\big)+X_t\Big)dt.
%&~~~~~~~~~
%-\frac{\beta+1}{\alpha}(u^-_tX_t)^\alpha\big(1-F_{u^-_tX_t}(u^-_tX_t)\big)^\beta+X_t\Big)dt.
\end{align*}

In accordance with Theorem $\ref{thm}$, its optimal solution $(\bar{u}_\cdot,\bar{X}_\cdot)$ should satisfy
\begin{equation}\label{p3}
p_t
=
\big(\bar{u}_t\bar{X}_t\big)^{\alpha-1}
\varpi_+'\big(1-F_{\bar{u}_t\bar{X}_t}(\bar{u}_t\bar{X}_t)\big), ~a.e. t\in[0,T], a.s.,
\end{equation}
where
\begin{equation*}
\left\{
\begin{array}{ccl}
dp_t&=&\big(\bar{u}_t p_t-q_t
-\big(\bar{u}_t\bar{X}_t\big)^{\alpha-1}
\varpi_+'\big(1-F_{\bar{u}_t\bar{X}_t}(\bar{u}_t\bar{X}_t)\big)\bar{u}_t
%&&+(\beta+1)\big(\bar{u}^-_t\bar{X}_t\big)^{\alpha-1}
%\big(1-F_{\bar{u}^-_t\bar{X}_t}(\bar{u}^-_t\bar{X}_t)\big)^\beta\bar{u}^-_t
-1 \big)dt+q_tdW_t,\\
p_T&=&0.
\end{array}
\right.
\end{equation*}
Combing these two equations, we have
\begin{equation*}
\left\{
\begin{array}{ccl}
dp_t&=&-(q_t+1)dt+q_tdW_t,\\
p_T&=&0.
\end{array}
\right.
\end{equation*}
It yields
\[
p_t=T-t,~q_t=0,~\forall t\in[0,T].
\]

Going back to equality $(\ref{p3})$, we write that $\bar{u}_t\bar{X}_t=h(p_t)$. If this is the case, $\bar{u}_t\bar{X}_t$ is deterministic and hence $F_{\bar{u}_t\bar{X}_t}(\bar{u}_t\bar{X}_t)=1$. As a result, we infer that
\[
\bar{u}_t\bar{X}_t=\Big(\frac{T-t}{(1-\nu)(\beta+1)}\Big)^{1/(\alpha-1)},~a.e.t\in[0,T], a.s..
\]
Substituting back to the state equation, we get
\[
\bar{X}_t=V_t\Big(x_0+\int^t_0  \Big(\frac{T-s}{(1-\nu)(\beta+1)V_s^{\alpha-1}}\Big)^{1/(\alpha-1)}   ds\Big),~~
V_t=\exp\big\{B_t-\frac{t}{2}\big\}.
\]
Finally, the optimal control is
\[
\bar{u}_t=\frac{(T-t)^{1/(\alpha-1)}}
{V_t\big(x_0((1-\nu)(\beta+1))^{1/(\alpha-1)}+\int^t_0  \frac{(T-s)^{1/(\alpha-1)}}{V_s}   ds\big)},~a.e.t\in[0,T], a.s..
\]
\end{example}

Without the distorted probability in this example, we acquire that
\[
\bar{u}_t=\frac{(T-t)^{1/(\alpha-1)}}
{V_t\big(x_0+\int^t_0  \frac{(T-s)^{1/(\alpha-1)}}{V_s}   ds\big)},~a.e.t\in[0,T], a.s..
\]

\section{\textsc{Concluding Remarks}}
This article develops a stochastic maximum principle for a general continuous behavioral portfolio model. The optimal solution is characterized by $(\ref{adjoint})$ and $(\ref{smp})$. The system $(\ref{state})$ and $(\ref{eqJ0})$ covers highly diversified preferences including those of the classical utility maximization, financial investment activities involving consumption (or gambling, insurance) and other behavioral patterns. Three examples are studied in last section, showing that our solution is in agreement with that of Jin and Zhou \cite{Jin and Zhou 2008}, the results are also used to solve optimization problems with distorted probabilities and running utilities.

Unlike the majority of models in literature, the running terms here are divided into positive and negative parts. The utility function is ill-behaved as a result of its $S$-shape and its infinite derivative at 0. Further, handling of  $F_Y(Y)$ on account of probability distortions poses serious mathematical challenges. To overcome these difficulties, we convert this setting to a mean-field optimal control problem, and derive a mean-field stochastic maximum principle. Due to a technical reason, we restricted  our utility as a one-variable function. We pose the study of the case when the utility function depends on more than one variable as a challenging open problem.

\addcontentsline{toc}{section}{Bibliography}

\end{document}